\documentclass[%
 aip,
 amsmath,amssymb,
 reprint,%
]{revtex4-1}

\usepackage{graphicx}
\usepackage{dcolumn}
\usepackage{bm}

\usepackage[utf8]{inputenc}
\usepackage[T1]{fontenc}
\usepackage{mathptmx}
\usepackage[dvipsnames]{xcolor}

\begin{document}

\preprint{AIP/123-QED}

\title{Dynamics of Ferroelectric Negative Capacitance - Electrostatic MEMS  Hybrid Actuators}

\author{Raghuram.T.R}
 \email{121704004@smail.iitpkd.ac.in}
\author{Arvind Ajoy}%
 \email{arvindajoy@iitpkd.ac.in}
\affiliation{ 
Department of Electrical Engineering, Indian Institute of Technology Palakkad, Palakkad, India
}%


\begin{abstract}
We propose a framework to model ferroelectric negative capacitance - electrostatic Micro Electro Mechanical Systems (MEMS) hybrid actuators and  analyze their dynamic (step input) response. Using this framework, we report the first proposal for reduction in the dynamic pull-in and pull-out voltages of the hybrid actuators due to the negative capacitance of the ferroelectric. The proposed model also reveals the effect of ferroelectric thickness on the dynamic pull-in and pull-out voltages and the effect of ferroelectric damping on the energy dissipated during actuation. We infer from our analysis that the hybrid actuators are better than the standalone MEMS actuators in terms of operating voltage and energy dissipation. Further, we show that one can trade-off a small part of the reduction in actuation voltage to achieve identical pull-in times in the hybrid and standalone MEMS actuators, while still consuming substantially lower energy in the former as compared to the latter. The circuit compatibility of the proposed hybrid actuator model makes it suitable for analysis and evaluation of various heterogeneous systems consisting of hybrid MEMS actuators and other electronic devices. 

\end{abstract}

\maketitle

\section{\label{sec:Introduction}Introduction}
Micro Electro Mechanical Systems (MEMS)  based  on electrostatic actuation  and sensing are  an integral part  of today's electronics. The ITRS roadmap describes  the significance of many such devices in applications ranging from consumer - to - automotive - to - medical   electronics \textcolor{blue}{\cite{itrs2015}}.   Electrostatic   MEMS actuators  are very  popular  and  are widely  used  because of  their inherent low  power consumption. The response of  such electrostatic MEMS actuators to voltage excitation is different for static and dynamic inputs.  The  former corresponds to the input  voltage being  varied slowly,  so that  the actuator  is in quasi-static equilibrium \textcolor{blue}{\cite{younis2011mems,elata2005static,lee2011principles}}. The latter  corresponds to the input voltage being  varied suddenly,  as in  the case  of a  step voltage excitation \textcolor{blue}{\cite{younis2011mems,elata2005static,lee2011principles,leus2008dynamic,nielson2006dynamic,rocha2004pull}}. It is well known that the dynamic response is different from the static response -- for example, the dynamic pull-in voltage of a MEMS cantilever is 0.919$\times$ its static pull-in voltage \textcolor{blue}{\cite{younis2011mems,nielson2006dynamic}}. 

The operating  voltage of  electrostatic MEMS actuators, for both static and dynamic inputs, is typically much  larger than the supply voltage  used in modern CMOS (Complementary Metal Oxide Semiconductor) integrated circuits: 10-100 V as compared to sub-1V CMOS circuit power supplies. 
\begin{figure}[b]
    \centering
    \includegraphics[scale=0.8]{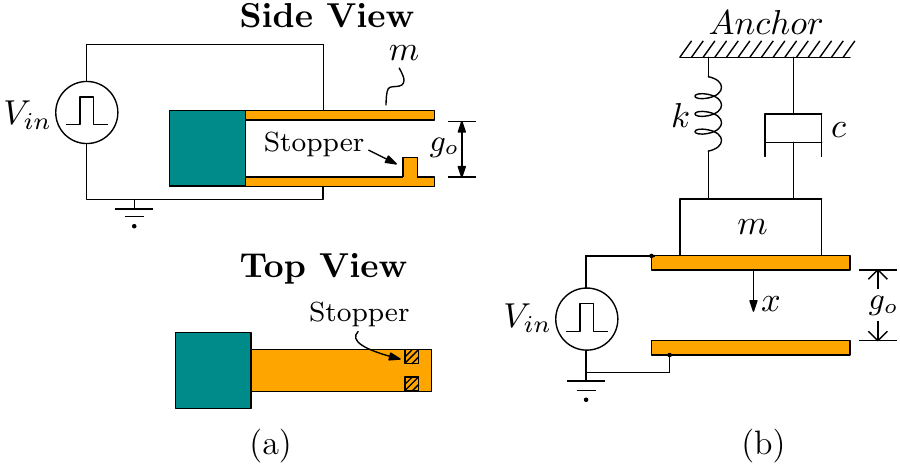}
    \caption{An  example   of  modeling a generic   electrostatic  MEMS
actuator using a 1-DOF model. (a) Cantilever MEMS (b) Equivalent 1-DOF
model capturing the essential features of the cantilever structure.}
    \label{fig:MEMS_1DOF}
\end{figure}
Additional on-chip voltage sources and drive electronics are used in present day MEMS-CMOS integrated circuits to meet the high operating voltage requirement of the MEMS actuators \textcolor{blue} {\cite{rebeiz2003rf,dumas2010design,saheb2007system,brandt2008high}}. Efforts to reduce the operating voltage also involve innovative designs like scaling down the structural parameters such as air-gap $<  $ 10  $nm$ \textcolor{blue}{\cite{lee2013sub,jang2008fabrication,attaran2015ultra}}. Reliable fabrication and operation of MEMS structures with extremely small air-gaps is very challenging due to stiction.

A novel technique to mitigate the demand for high operating voltage of an electrostatic MEMS actuator was proposed in Ref. \textcolor{blue}{\onlinecite{masuduzzaman2014effective}}, by connecting a ferroelectric capacitor exhibiting negative capacitance in series with the MEMS actuator, thus forming a hybrid actuator. The "static response" of this hybrid actuator was analyzed in Refs. \textcolor{blue}{\onlinecite{masuduzzaman2014effective}} and \textcolor{blue}{\onlinecite{choe2017adjusting}}, where the operating voltage to static inputs was analytically proven to be lowered in the hybrid actuator as compared to the standalone MEMS actuator. The impact of ferroelectric negative capacitance on the energy-delay characteristics of the actuator was theoretically predicted in Ref. \textcolor{blue}{\onlinecite{choe2019impact}} -- however important effects such as the ferroelectric switching delay and the dependence of electrostatic force on  displacement have been ignored therein.  

The dynamic response of electrostatic MEMS actuators to step inputs play  a  key  role in  switching  (like RF  MEMS switches) and display applications \textcolor{blue}{\cite{nielson2007dynamic,shekhar2011switching, rebeiz2004rf}}. The dynamic response of hybrid actuators has not been analyzed in the literature. Our work deals with (a) numerical modeling of the hybrid actuator by solving the non-linear differential equations governing its dynamics, (b) investigating its "dynamic response" to a step input, (c) analyzing the effect of ferroelectric parameters on the dynamic response, (d) studying the trade-off  between pull-in time and  operating voltage and its implication on energy dissipation in the hybrid actuator.  
We use SPICE (Simulation Program with Integrated Circuit Emphasis) to solve the differential equations mentioned above. SPICE is a popular tool for circuit analysis. Hence our approach leads to a model that can be included seamlessly to evaluate the performance of  CMOS-MEMS hybrid  circuits. 

This paper is organized as follows. Section \textcolor{blue}{\ref{sec:Standalone_MEMS}} reviews the dynamics of
a standalone  electrostatic  MEMS  actuator. Section  \textcolor{blue}{\ref{sec:Model}}  presents  the
modeling of  the ferroelectric negative  capacitance -  electrostatic MEMS
hybrid  actuator. Simulation  results and  discussion are
detailed in section \textcolor{blue}{\ref{sec:Results}}. Section \textcolor{blue}{\ref{sec:Conclusion}} presents our conclusion.
\begin{figure*}[t]
    \centering
    \includegraphics[scale=0.8]{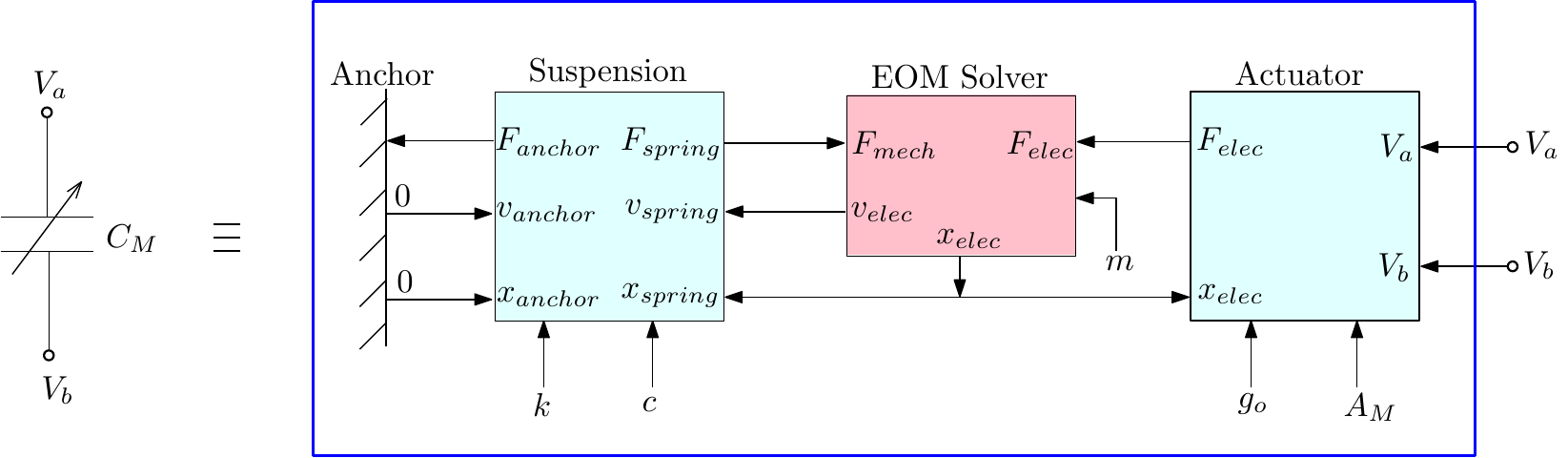}
    \caption{SPICE    model   of    standalone   electrostatic    MEMS
actuator.  Arrows pointing  into  (out  of) a  block  refer to  inputs
(outputs) to (from) the block. $x$ designates position, $v$ designates
velocity  and $F$  designates forces.  Implementation of  these blocks
using circuit elements follows Ref. \textcolor{blue}{\onlinecite{toshiyoshi2011spice}}. Refer \textcolor{blue}{supplementary material} for detailed schematic implementation.} 
    \label{fig:MEMS_SPICE}
\end{figure*}
\section{\label{sec:Standalone_MEMS}Review of Dynamics of Electrostatic MEMS Actuator}
In  this  work,  we  model a  generic  standalone  electrostatic  MEMS
actuator  as  a  single  degree of  freedom  (1-DOF)   parallel  plate
arrangement consisting of a pair  of electrodes separated by an air-gap
$g_o$. As shown in the Fig. \textcolor{blue}{\ref{fig:MEMS_1DOF}}, one electrode (bottom) is fixed and the
other (top) is movable. The parameters  used in the 1- DOF model (mass
$m$,  spring constant  $k$, damping  coefficient $c$)  represent their
effective equivalents  in the generic  actuator. We set
the damping  coefficient to zero,  so as  to enable a  comparison with
analytical results  wherever possible.  For dynamic  operation, a step
voltage is  applied to the  actuator.  Below  a certain value  of this
applied step voltage, called the dynamic pull-in voltage, the response
of  the actuator  is periodic.   The  maximum value  of this  periodic
displacement of the electrode  is called dynamic pull-in displacement.
When  the applied  step voltage  is larger than the  dynamic pull-in  voltage,  the
movable  top electrode  snaps down  onto the  bottom electrode.   This
condition  is called  dynamic  pull-in \textcolor{blue}{\cite{leus2008dynamic}}.   After
achieving pull-in,  when the  applied step voltage  is decreased  to a
specific value, called the dynamic release voltage / pull-out voltage,
the pull-in condition is lost and  the movable top electrode gets detached from the fixed bottom electrode. This condition is termed as release / pull-out. We assume that  the MEMS actuator has a pair of
stoppers with height $h_s$, as shown  in Fig.  \textcolor{blue}{\ref{fig:MEMS_1DOF}(a)}, to minimize the  area of contact upon
pull-in  and thus  minimize  the effect of surface forces.  We hence
neglect these forces in our analysis.

The equation of motion governing the actuator response to the applied
voltage $V_{in}~u(t)$, where $u(t)$ is the unit step function, is given by
\begin{align} \label{eq:dynamics_eom}
m~\frac{d^2x}{dt^2} + c~\frac{dx}{dt} + k~x = \frac{\epsilon_o~A_M~V_{in}^2}{2~(g_o-x)^2}
\end{align}
for $t \ge 0$. $A_M$ is the area of the electrode, $\epsilon_o$ is the
permittivity  of  free   space,  and  $x$  is   the  dynamic  variable
representing  the   displacement  of   the  electrode.   With  damping
neglected,  the  dynamic  pull-in  voltage $V_{DPI}$ , the  dynamic
pull-in displacement $X_{DPI}$ and pull-out voltage $V_{DPO}$ are given by \textcolor{blue}{\cite{lee2011principles}}
\begin{align} \label{eq:pull_in_formula}
V_{DPI} = \sqrt{\frac{k g_o^3}{4 \epsilon_o A_M}}~;~X_{DPI} = \frac{g_o}{2} ~;~ V_{DPO}=\sqrt{\frac{2 k h_s^2 (g_o-h_s)}{\epsilon_o A_M}}
\end{align}
As a reference, the respective static pull-in quantities are \textcolor{blue}{\cite{younis2011mems}}
\begin{align} \label{eq:static_pull_in_formula}
V_{SPI} = \sqrt{\frac{ 8 k g_o^3}{27 \epsilon_o A_M}}~;~X_{SPI} = \frac{g_o}{3} ~;~ V_{SPO}= V_{DPO}
\end{align}
\section{\label{sec:Model}Modeling Ferroelectric Negative Capacitance - Electrostatic MEMS Hybrid Actuator}
SPICE is a general purpose circuit simulation program for nonlinear DC, nonlinear transient and linear AC analysis. SPICE \textcolor{blue}{\cite{vladimirescu1994spice}} is used as a modeling program to mathematically predict the behavior of electronic circuits. Although SPICE was originally developed for electronic circuit simulation, it has been extended for design and analysis of problems in various areas of physics -- thermal, electro-thermal \textcolor{blue}{\cite{vogelsong1989extending,chavez2000spice}}, optics \textcolor{blue}{\cite{xu1997oe,ravezzi2000versatile}}, mechanics \textcolor{blue}{\cite{leach1994controlled}}, biological \textcolor{blue}{\cite{madec2017modeling}} and microfluidics \textcolor{blue}{\cite{takao2011micro}}. Here we use SPICE to numerically solve the differential equations governing the actuator dynamics. This involves using arbitrary voltage sources, current sources and built-in integrate and differentiate functions available in SPICE. Using SPICE to model the actuator is advantageous as SPICE is circuit compatible and thus, the model can be used with other electronic devices to evaluate various heterogeneous systems for different applications.   

The SPICE model of the standalone electrostatic MEMS is implemented as
shown in Fig. \textcolor{blue}{\ref{fig:MEMS_SPICE}}, based on Ref. \textcolor{blue}{\onlinecite{toshiyoshi2011spice}}. It consists  of four modules namely the actuator, the suspension,  the  Equation  Of  Motion (EOM)  solver  and  the  anchor. These   modules  are   represented   as
sub-circuits in  the schematic along with  their associated parameters
as depicted in Fig. \textcolor{blue}{\ref{fig:MEMS_SPICE}}. See \textcolor{blue}{supplementary material} for implementation details.  The  initial displacement
and velocity of the movable electrode  are taken as zero. The actuator
module takes  the applied step  voltage $V_{in}$, area  $A_M$, initial
air-gap  $g_o$  and the  electrode  displacement  $x_{elec}$ as  input
parameters  and   calculates  the  electrostatic  force   $F_{elec}  =
{\epsilon_o A_M  V_{in}^2}/{2(g_o-x)^2}$. The suspension  module takes
the electrode displacement  $x_{elec}$, electrode velocity $v_{elec}$,
the spring constant  $k$ and the damping coefficient $c$  as the input
parameters and calculates the mechanical restoring force $F_{mech} = c
\cdot {dx}/{dt}  + k x$. The  EOM solver module is  placed between the
actuator  and suspension  modules.  The EOM  solver  compares the  two
forces  $F_{elec}$   and  $F_{mech}$  and  calculates   the  acceleration as $(F_{elec}-F_{mech})/m$ according to Eq.\textcolor{blue}{(\ref{eq:dynamics_eom})} . The acceleration is then integrated to obtain the velocity, which is further integrated to obtain the displacement. Feedback  connections  are
provided within these blocks to  determine the electrode  displacement  for  an  applied  voltage. Note that non-electrical quantities are modeled using currents or voltages -- suitable conversion factors are therefore required in order to read the final results with appropriate units.
\begin{figure}[t]
    \centering
    \includegraphics[scale=0.9]{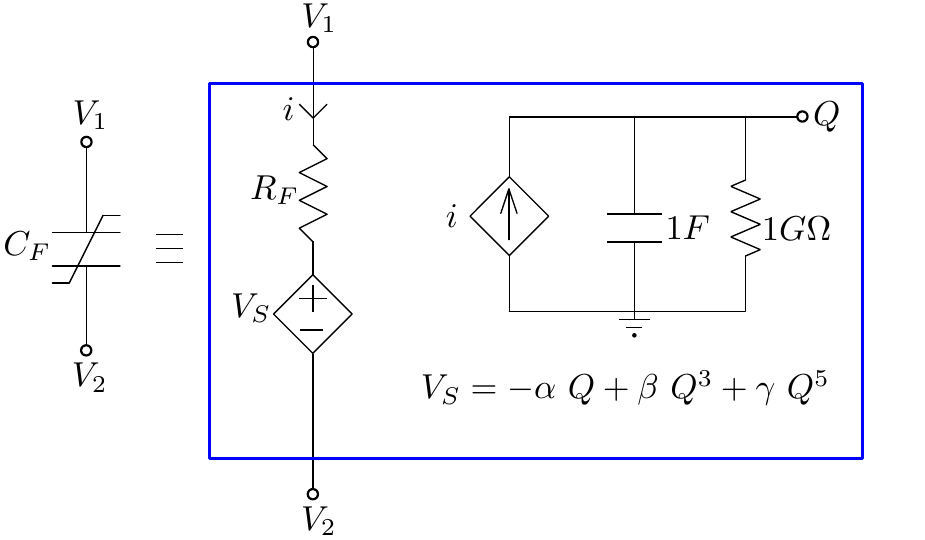}
    \caption{SPICE model of ferroelectric capacitor. The charge $Q$ is
determined  using  an  RC   integrator. The  voltage  source  $V_S$  is
implemented using arbitrary behavioral voltage source in SPICE.}
    \label{fig:FERRO_SPICE}
\end{figure}

The  dynamics  of  the  ferroelectric  capacitor  (single  domain)  is
captured  by the  time dependent  Landau -  Khalatnikov (LK)  equation
\textcolor{blue}{\cite{salahuddin2008use,khan2015negative,aziz2016physics,li2017delay,
li2017evaluation,khan2015negative_thesis}}  relating the  voltage $V_F$
across the ferroelectric to charge $Q$ as
\begin{align} \label{eq:ferro_volt}
V_F &= -\alpha~Q + \beta~Q^3 + \gamma~Q^5 + R_F~\frac{dQ}{dt}
\end{align}
\begin{align} \label{eq:ferro_coeff}
  \alpha = -\frac{\alpha_F~t_F}{A_F} ;~
  \beta = \frac{\beta_F~t_F}{A_F^3}  ;~
  \gamma = \frac{\gamma_F~t_F}{A_F^5} ;~
  R_F = \frac{\rho~t_F}{A_F}
\end{align}
where  $\rho$  is  the  ferroelectric  damping  constant;  $\alpha_F$,
$\beta_F$  and $\gamma_F$  are ferroelectric  anisotropy coefficients,
$t_F$, $A_F$  are respectively the  thickness and area of  the ferroelectric.  The last  term of
Eq.\textcolor{blue}{(\ref{eq:ferro_volt})}  denotes the voltage  drop across resistor $R_F$  with $dQ/dt$
representing the current $i$ through it.  Thus, Eq. \textcolor{blue}{(\ref{eq:ferro_volt})} is implemented
in SPICE  as a Voltage Controlled  Voltage Source (VCVS) in  series with
resistor $R_F$ \textcolor{blue}{\cite{li2017delay}}.   The SPICE  model of  the ferroelectric  capacitor is
shown  in Fig.  \textcolor{blue}{\ref{fig:FERRO_SPICE}}.  The  charge $Q$  is estimated  by integrating  
current          $i$           through          the          capacitor
\textcolor{blue}{\cite{pevsic2017computational}}. 
 
\begin{figure}[t]
\vspace{0.30in}
    \centering
    \includegraphics[scale=1.0]{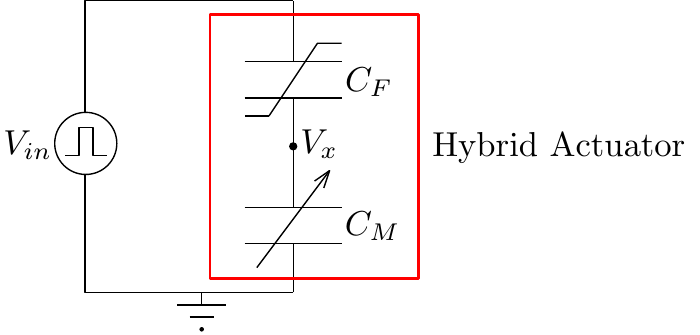}
    \caption{Ferroelectric  negative  capacitance -electrostatic  MEMS
hybrid actuator equivalent circuit. $C_F$ represents the ferroelectric
capacitor and  $C_M$ represents  the variable capacitance  provided by
the MEMS actuator.}
    \label{fig:NC_MEMS_EqCkt}
\end{figure}

We propose a SPICE model for the  ferroelectric negative  capacitance -  electrostatic MEMS  hybrid
actuator   by cascading the sub-circuit corresponding to the  ferroelectric capacitor
$C_F$  with that of the standalone  MEMS  actuator  (depicted as  a  variable
capacitor $C_M$)  as shown in  Fig. \textcolor{blue}{\ref{fig:NC_MEMS_EqCkt}}. This solves the differential equations Eqs.\textcolor{blue}{(\ref{eq:dynamics_eom}, \ref{eq:ferro_volt})}, by ensuring that an identical charge exists on the ferroelectric capacitor and the MEMS actuator.
\section{\label{sec:Results}Simulation Results and Discussion}
Table  \textcolor{blue}{\ref{tab:Table1}} lists  the  parameters of  the hybrid  actuator  used in  the
simulation.   SBT
(Sr$_{0.8}$Bi$_{2.2}$Ta$_2$O$_9$)    is    chosen   as    the    ferroelectric
material. The ferroelectric layer thickness and area are designed so
as to  obtain a static pull-in  voltage of $0.80~V$ and  a static pull-out
voltage   of    $0.00~V$,   using   the   equations    available   in   Ref.
\textcolor{blue}{\onlinecite{masuduzzaman2014effective}}.  The validation  of the  SPICE model
for  static  pull-in and  pull-out analysis  of the hybrid actuator can be  found in  the
\textcolor{blue}{Appendix}. As mentioned  earlier, the movement of the  top electrode is
limited by means of the stopper of height $h_s$.
\subsection{Dynamic response of the actuator}
\begin{figure}[t]
    \centering
    \includegraphics[scale=0.75]{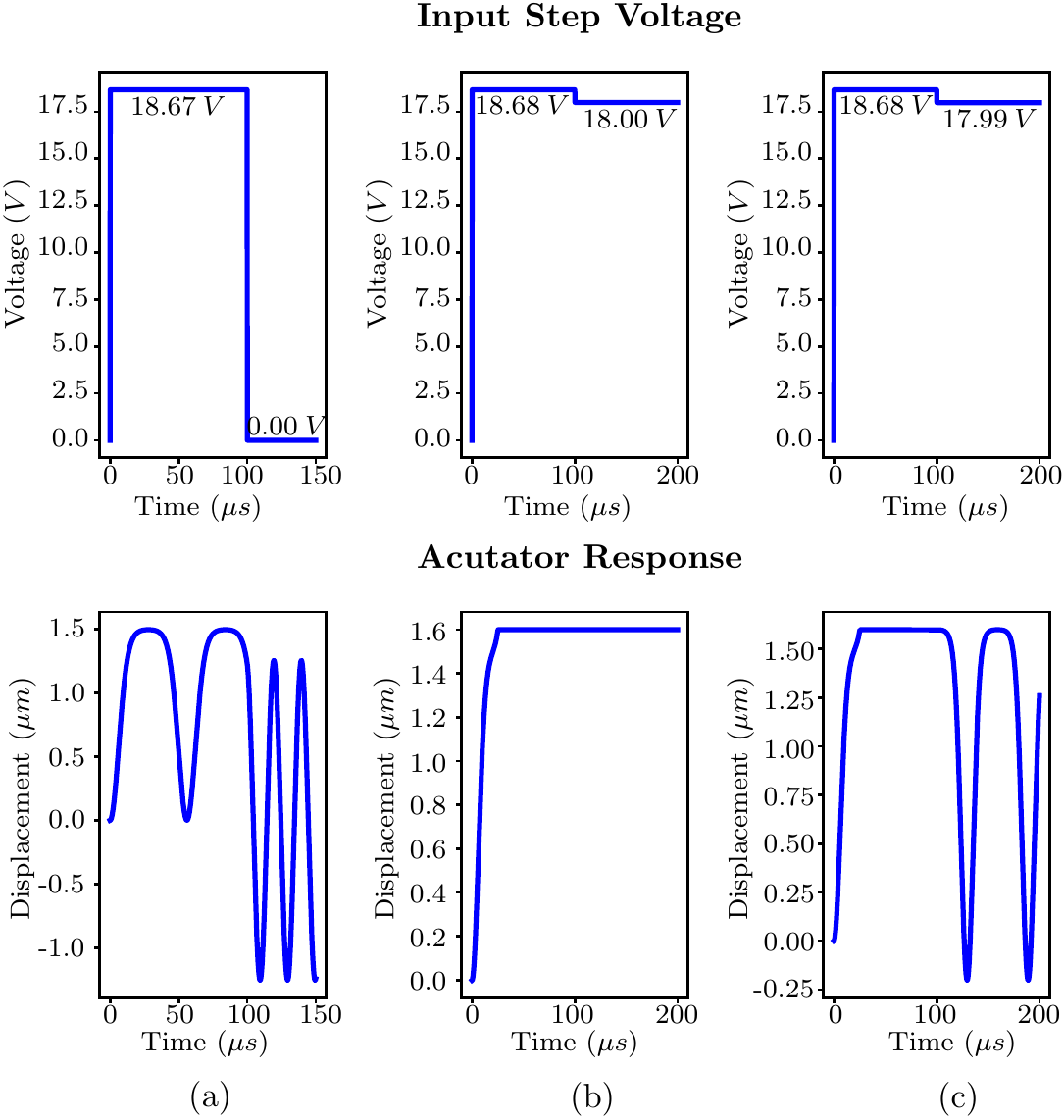}
    \caption{Standalone electrostatic  MEMS actuator  dynamic response
(a)  Actuator  response  before  dynamic  pull-in,  showing  dynamic pull-in displacement $X_{DPI}$ = 1.50  $\mu  m$. (b)  Actuator  response after  dynamic
pull-in  and  without release.  Note that the stopper restricts the displacement to 1.6 $\mu m$. (c)  Actuator  response after  dynamic
pull-in and with release. Note that the pull-in voltage $V_{DPI}$ is 18.67 $V$ and release voltage $V_{DPO}$ is 17.99 $V$.}
    \label{fig:Dynamic_Standalone_MEMS_Resp}
\end{figure}

Fig. \textcolor{blue}{\ref{fig:Dynamic_Standalone_MEMS_Resp}}  shows the  simulated dynamic response  of the  standalone MEMS
actuator.  We obtain a  dynamic pull-in
voltage $V_{DPI}$ of $18.67~V$, a pull-in displacement $X_{DPI}$ of $1.50~\mu m$ and pull-out voltage $V_{DPO}$ of $17.99~V$ from the simulation.  
\begin{figure}[t]
    \centering
    \includegraphics[scale=0.75]{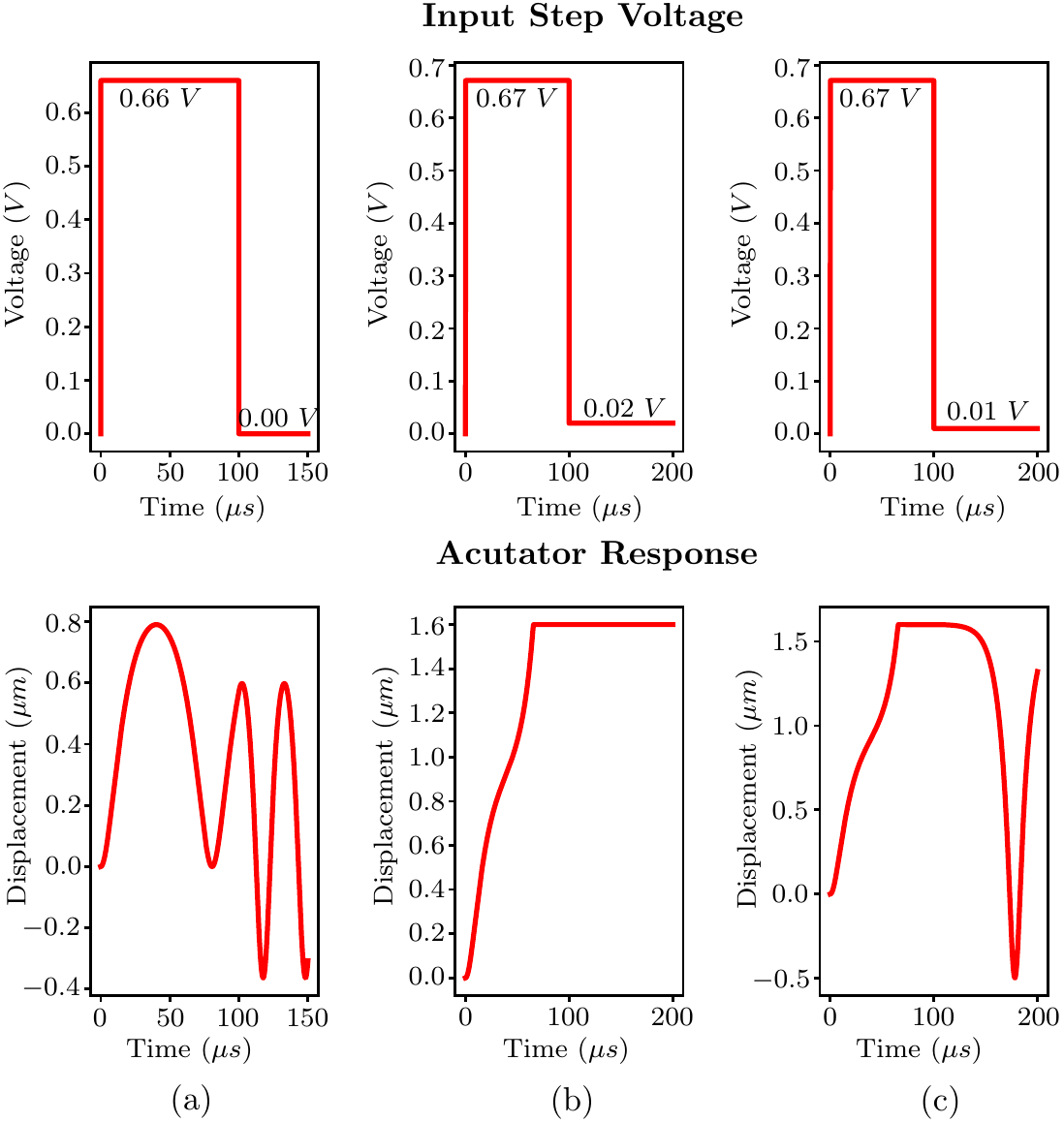}
    \caption{Ferroelectric negative  capacitance -  electrostatic MEMS
hybrid  actuator dynamic  response  showing  significant reduction  in
operating voltage. (a) Actuator  response before dynamic pull-in, showing a reduced dynamic pull-in displacement $X_{HDPI}=0.79~\mu m$. (b)
Actuator  response  after dynamic  pull-in  and  without release.  (c)
Actuator response after dynamic pull-in and with release. Note that the pull-in voltage is reduced to $V_{HDPI}$ = 0.66 $V$ and release voltage is reduced to $V_{HDPO}$ = 0.01 $V$.}
    \label{fig:Dynamic_NC_MEMS_Resp}
\end{figure}
\begin{table}[t]
\caption{\label{tab:Table1}%
Ferroelectric negative capacitance - electrostatic MEMS hybrid actuator parameters used for SPICE simulation
}
\begin{ruledtabular}
\begin{tabular}{ll}
\textrm{Parameter}&
\textrm{Value}\\
\colrule
Length of the cantilever, $L$ & $160 ~\mu m$ \\
Width of the cantilever, $W$ & $6 ~\mu m$ \\
Thickness of the cantilever, $T$ & $2 ~\mu m$ \\
Cantilever Material & Silicon (Si)\\
Young's Modulus, $E$ & $150 ~GPa$~\textcolor{blue}{\cite{kim2001111}} \\
Density, $D$ & $2330 ~kg/m^3$~\textcolor{blue}{\cite{shackelford2016crc}} \\
Mass, $m$ & $4.4736 ~X~ 10^{-12}  ~kg$\\
Spring Constant, $k$ & 0.439 $N/m$\\
Initial air-gap, $g_o$ & $3~\mu m$\\
Stopper height, $h_s$ & $1.4~\mu m$\\
Permittivity of free space, $\epsilon_o$ & $8.854 ~X~ 10^{-12}~F/m$\\
Ferroelectric material & $SBT (Sr_{0.8}Bi_{2.2}Ta_2O_9)~\textcolor{blue}{\cite{masuduzzaman2014effective}}$\\
$\alpha_F$ & $-6.5 ~X~ 10^{7}~m/F$\\
$\beta_F$ & $3.75 ~X~ 10^{9}~m^5/F/C^2$\\
$\gamma_F$ & $0~m^9/F/C^4$\\
Ferroelectric thickness, $t_F$ & $5.99~\mu m$\\
Ferroelectric area, $A_F$ & $1.1659~X~10^{-12}~m^2$\\
\end{tabular}
\end{ruledtabular}
\end{table}
Our simulation   results    exactly match with    the   analytical predictions given by Eq. \textcolor{blue}{(\ref{eq:pull_in_formula})}.  For input  step voltage  less than  the dynamic  pull-in voltage (18.67  $V$), the  actuator response  is periodic.   For input step voltage  greater than the  dynamic pull-in voltage,  the actuator achieves  pull-in. After pull-out, the top electrode oscillates, in the absence of damping.

\begin{figure}[b!]
    \centering
    \includegraphics[scale=0.90]{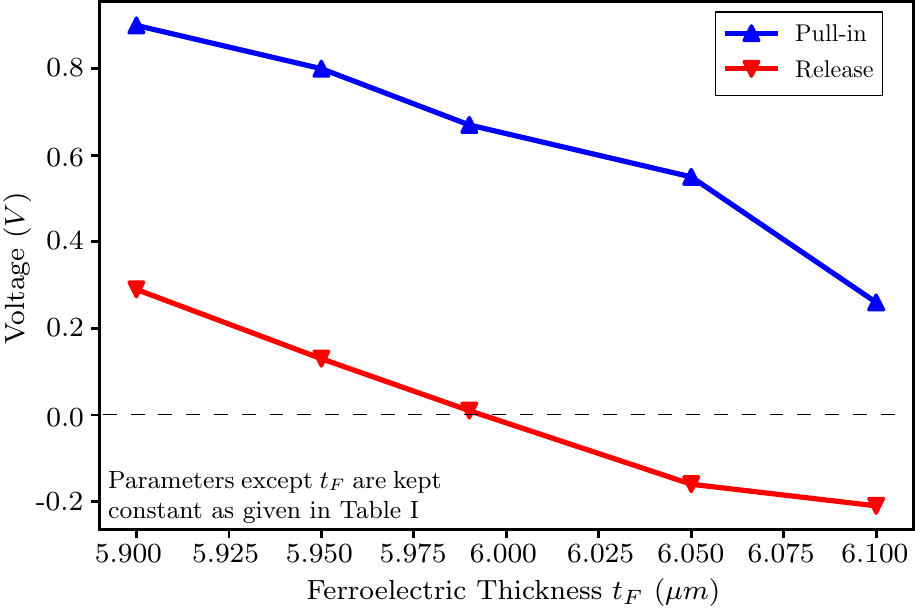}
    \caption{Effect of ferroelectric thickness $t_F$ on the dynamic pull-in and release voltages of the hybrid actuator depicting the possibility of obtaining both positive and negative release voltages by proper choice of ferroelectric thickness.}
    \label{fig:Dynamic_NC_MEMS_Thcikness}
\end{figure}  
The ferroelectric  capacitor model is  now connected in series  with the
standalone MEMS model to form the hybrid actuator.  Again, the dynamic
response of the  hybrid actuator (with $\rho$ = 0) is obtained by applying  a step input
voltage as shown in Fig. \textcolor{blue}{\ref{fig:Dynamic_NC_MEMS_Resp}}. For the hybrid actuator, the simulation results give a dynamic  pull-in voltage $V_{HDPI}$ of 0.66 $V$, pull-in displacement $X_{HDPI}$ of 0.79 $\mu m$ and release  voltage $V_{HDPO}$ of 0.01 $V$. Compared to  the
standalone  MEMS actuator,  there is  a significant  reduction in  the
operating  voltage of  the hybrid  actuator. This  is because  of the
voltage amplification  due to the  negative capacitance of  the series
ferroelectric capacitor. This sub-1V  operation of the hybrid actuator
should allow seamless  integration of such MEMS  actuators with modern
CMOS  devices,  eliminating  the  need for  any  drive  electronics  or
additional  on-chip voltage  up-converters. Simulation using SPICE  enables a view of the time history response (input vs. time, displacement vs. time) distinctly before \& after pull-in and before \& after pull-out as shown in Figs. \textcolor{blue}{\ref{fig:Dynamic_Standalone_MEMS_Resp}} and \textcolor{blue}{\ref{fig:Dynamic_NC_MEMS_Resp}}.
\begin{figure}[t]
	\centering
    \includegraphics[scale=0.70]{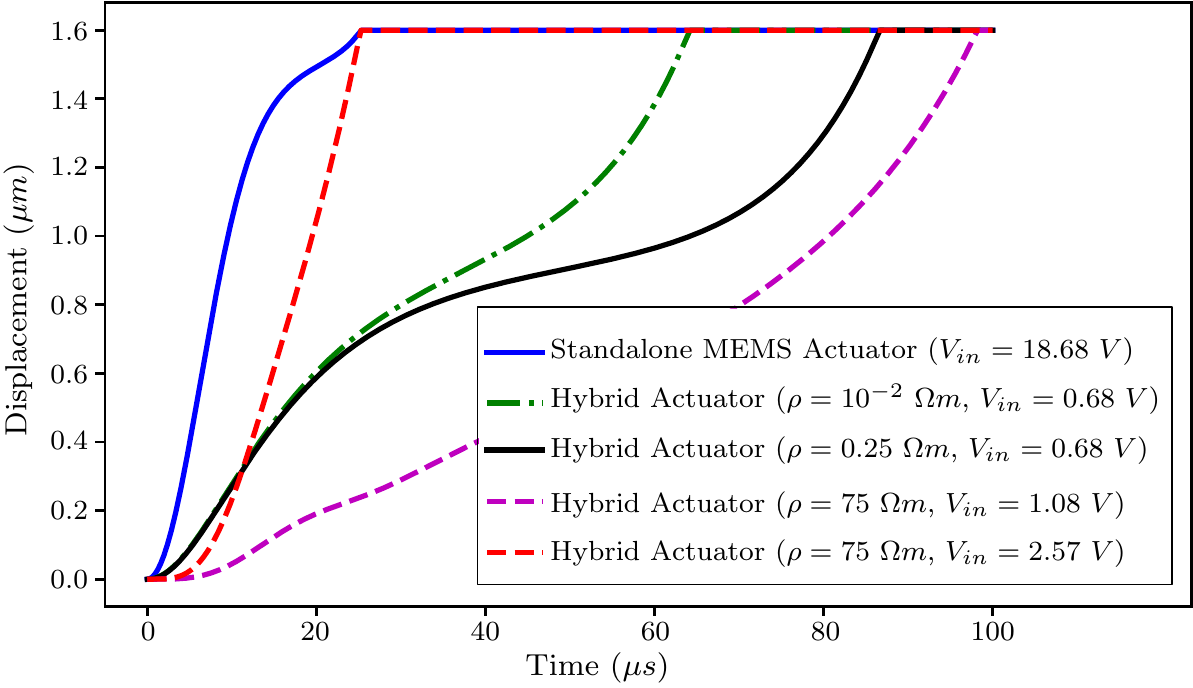}
    \caption{Dynamic pull-in time analysis of standalone MEMS actuator and hybrid actuator (for different $\rho$). It is shown that the hybrid actuator is slower in comparison with the standalone MEMS actuator to achieve dynamic pull-in. There also  exists a trade-off between the applied step voltage and the pull-in time.}
    \label{fig:Dynamic_NC_MEMS_Time_Analysis}
\end{figure}
\subsection{Effect of ferroelectric thickness on operating voltage}
The effect of ferroelectric thickness $t_F$ on the dynamic pull-in and
release voltages  of the hybrid actuator (with $\rho$ = 0) is shown  in Fig. \textcolor{blue}{\ref{fig:Dynamic_NC_MEMS_Thcikness}}.   Increase in  the ferroelectric
thickness  reduces  the  operating  voltage. This  is  because,  with
increase  in ferroelectric  thickness,  the ferroelectric  capacitance
decreases  leading  to enhanced  voltage  amplification.   This is  in
agreement  with a  similar  trend observed  in ferroelectric  negative
capacitance -  FET (Field Effect Transistor) devices  \textcolor{blue}{\cite{li2017delay}} where the  gate voltage
reduces  with  increase  in  ferroelectric  thickness.   However,  the
ferroelectric  thickness should  be properly  chosen (i.e.  $|C_F| \sim
C_M$) so  as to preserve  the voltage amplification  phenomenon.  Note
that the release voltage can  be either positive or negative. Negative
release  voltage  is  favorable  for  bipolar  voltage  actuation  of
electrostatic   MEMS  actuators   leading   to  improved   reliability
\textcolor{blue}{\cite{peng2007dielectric,dominguez2010dielectric}}  and also  in memory
applications \textcolor{blue}{\cite{choi2008nano}}. However, this requires both positive
and negative  power supplies  for operation.   Tailoring ferroelectric
thickness $t_F$ can also ensure 0 $V$ release facilitating the
use of  single sub-1V voltage  source for both pull-in  and release.
For  example, a  hybrid actuator  with $t_F  = $  6.10 $\mu  m$ has  a
negative release voltage  (-0.21 $V$), which implies  that reducing the  applied  voltage to  0  $V$ will  not  result in  release.
However, a  small change in  thickness to  5.95 $\mu m$  (with release
voltage 0.13 $V$) will ensure release of the cantilever at 0 $V$.
\begin{figure}[b]
    \centering
    \includegraphics[scale=0.90]{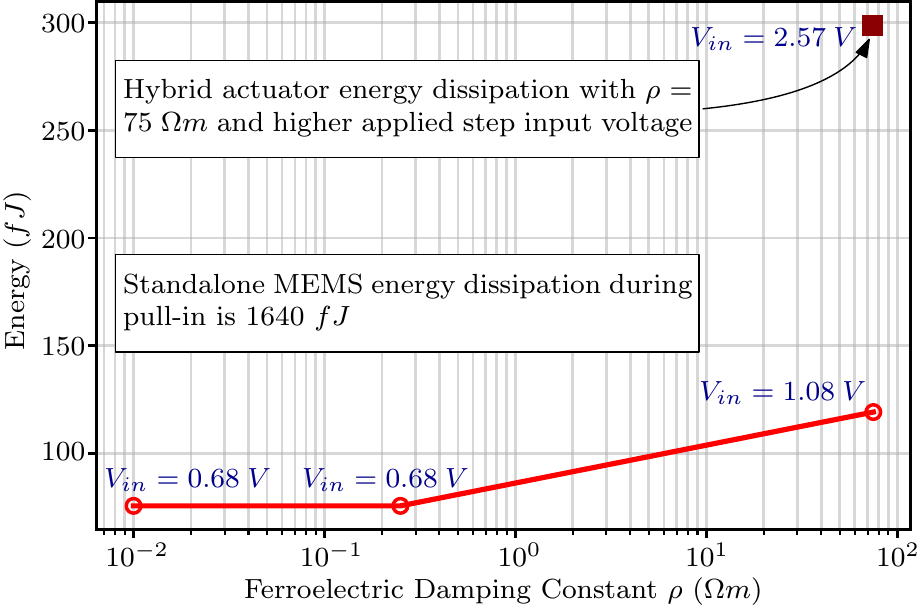}
    \caption{Effect of ferroelectric damping constant $\rho$ on the energy dissipation during dynamic pull-in in the hybrid actuator. Note that there is reduction in the energy dissipation in comparison with the standalone MEMS actuator due to reduction in the dynamic pull-in voltage.}
    \label{fig:Dynamic_NC_MEMS_Energy}
\end{figure}
\subsection{Temporal analysis of dynamic pull-in}
A temporal analysis of dynamic pull-in of the standalone MEMS actuator
and the hybrid actuator is shown in Fig. \textcolor{blue}{\ref{fig:Dynamic_NC_MEMS_Time_Analysis}}. The release process is not
considered as  displacement by  only few  nanometers is  sufficient to
release  the   electrode.   Ferroelectric  damping,  modeled   by  the
resistance  $R_F  =  \rho  ~t_F  /  A_F$ in  the  SPICE  model  of  the
ferroelectric capacitor, plays an important  role in the time response
of  the hybrid  actuator.  A  large  range of  $\rho$ is  used in  the
simulation  for a  comprehensive  prognosis. The pull-in voltage  is a  function of $\rho$.  For each value of  the ferroelectric damping constant, a step input voltage 10 $mV$  greater than the pull-in voltage,  is used for actuation.   It is  observed that  the  hybrid actuator  is slower  in comparison with the standalone MEMS  actuator. However, note that there is a trade-off between the applied  step input voltage and the pull-in time -- a  larger step input voltage will result  in faster pull-in of the actuator.  This  suggests that, by applying a  higher step voltage (which is  still smaller  than that of  the standalone  actuator), the pull-in time of  the hybrid actuator can be made  equal to the pull-in
time of  the standalone actuator.  This  ensures low-voltage operation
of the hybrid actuator without  compromising on the pull-in time.  For
example, both standalone MEMS actuator and hybrid actuator (with $\rho$
= 75 $\Omega  m$ and step input  voltage $V_{in}$ = 2.57  $V$) have the
same dynamic pull-in time as depicted in Fig. \textcolor{blue}{\ref{fig:Dynamic_NC_MEMS_Time_Analysis}}.
\subsection{Effect of ferroelectric damping on energy dissipation}
Electrostatic MEMS actuators are inherently low-power devices owing to
their          near           zero          power          dissipation
\textcolor{blue}{\cite{brown1998rf,rebeiz2001rf}}. Reduction  in the  operating voltage  can further reduce  the power dissipated during switching and thus enable the use  of such actuators in low-power, low-voltage applications. We estimate  the energy for pull-in as the time integral of the  instantaneous power  $v_{in}(t) \cdot  i(t)$ over  the entire  pull-in time.   Fig. \textcolor{blue}{\ref{fig:Dynamic_NC_MEMS_Energy}}  shows the  results  of the  above calculation.  Again, ferroelectric damping constant  $\rho$ plays an important  role. As in the  case of  temporal  analysis, we  apply a  step  voltage 10 $mV$ larger than  the  pull-in  voltage in  each  case to  ensure dynamic pull-in.  Note  that  the  energy dissipated  during  pull-in  in  the
standalone MEMS actuator is 1640  $fJ$.  The energy dissipated during
pull-in in the  hybrid actuator for different values  of $\rho$ varies
between  50-150 $fJ$.  For example,  for $\rho$  = 75  $\Omega~m$, the
energy dissipated  during pull-in  is 119.17 $fJ$.   This shows  a 10x
reduction    in   the    energy   dissipation    using   the    hybrid
actuator.  However, this  reduction in  energy  comes at  the cost  of
slower pull-in. In the previous paragraph, we showed that the actuator
can be operated using a voltage higher than pull-in voltage to achieve
a   pull-in  time   identical   to  that   of   the  standalone   MEMS
actuator.  Fig. \textcolor{blue}{\ref{fig:Dynamic_NC_MEMS_Energy}}  also shows  the energy  dissipated when  the hybrid
actuator with $\rho$  = 75 $\Omega m$ is actuated  with a step voltage
$V_{in}$ = 2.57  $V$.  We find that the energy  dissipated is still 5x
lower  than  the energy dissipated in the standalone  actuator,  indicating  a very  favorable
application of the trade-off between pull-in time and applied voltage.
\section{\label{sec:Conclusion}Conclusion}
We presented a SPICE based framework to model ferroelectric negative capacitance - electrostatic MEMS hybrid actuator and to analyze the dynamic (step input) response of the hybrid actuator. It is shown that the  dynamic pull-in and release voltages of  this hybrid actuator are  significantly reduced due to the  presence of the  series ferroelectric capacitor exhibiting negative capacitance. This allows straightforward  integration  of  such   actuators  with  modern  CMOS devices.   Further, this  also  opens the  door for  the  use of  such actuators  in  low-power,  low-voltage switching applications.   The  effect  of ferroelectric  thickness  in  achieving  both  positive  and  negative release voltage  is also  illustrated. During dynamic pull-in, there is  considerable reduction  in the energy  dissipated  in the  hybrid  actuator  as compared  to  the standalone MEMS actuator, accompanied however by an increase in pull-in  time.  Nevertheless, we  can  trade-off pull-in  time  versus  applied voltage  to  achieve identical pull-in times for the hybrid and standalone actuators and still achieve reduction in the energy dissipated in the hybrid actuator.   Finally, since the proposed model is SPICE
based and  thus circuit  compatible, this can  be used  in combination
with other low-voltage CMOS  circuits to analyze various heterogeneous CMOS - MEMS systems.
\begin{acknowledgments}
The authors thank Dr. Revathy  Padmanabhan, Dr. Sukomal Dey, Indian
Institute of Technology Palakkad, Palakkad and Prof. G. K. Ananthasuresh, Indian Institute of Science, Bengaluru for helpful discussions.
\end{acknowledgments}

\appendix*
\section{Static characteristics of the hybrid actuator}
The static pull-in voltage $V_{HSPI}$ and the travel range $X_{HSPI}$ of the hybrid actuator are given by \textcolor{blue}{\cite{masuduzzaman2014effective}}
\begin{equation} \label{eq:hybrid_static} \tag{A1}
V_{HSPI} = r_{\alpha N}~ \sqrt{\frac{r_{\alpha N}}{r_{\beta N}} \cdot \frac{8~k~g_o^3}{27~\epsilon_0~A_M}}; ~X_{HSPI} = \frac{r_{\alpha N}}{r_{\beta N}} \cdot \frac{g_o}{3}
\end{equation}
where
\begin{equation} \tag{A2}
r_{\alpha N} = 1- {\frac{t_F~A_M~|\alpha_F|~\epsilon_o}{g_o~A_F}} 
\end{equation}
\begin{equation} \tag{A3}
r_{\beta N} = 1- \left[(2~\beta_F~k~\epsilon_o^2)\left(\frac{t_F~A_M^2}{A_F^3}\right)\right]
\end{equation} 
\begin{figure}[t]
    \centering
    \includegraphics[scale=0.70]{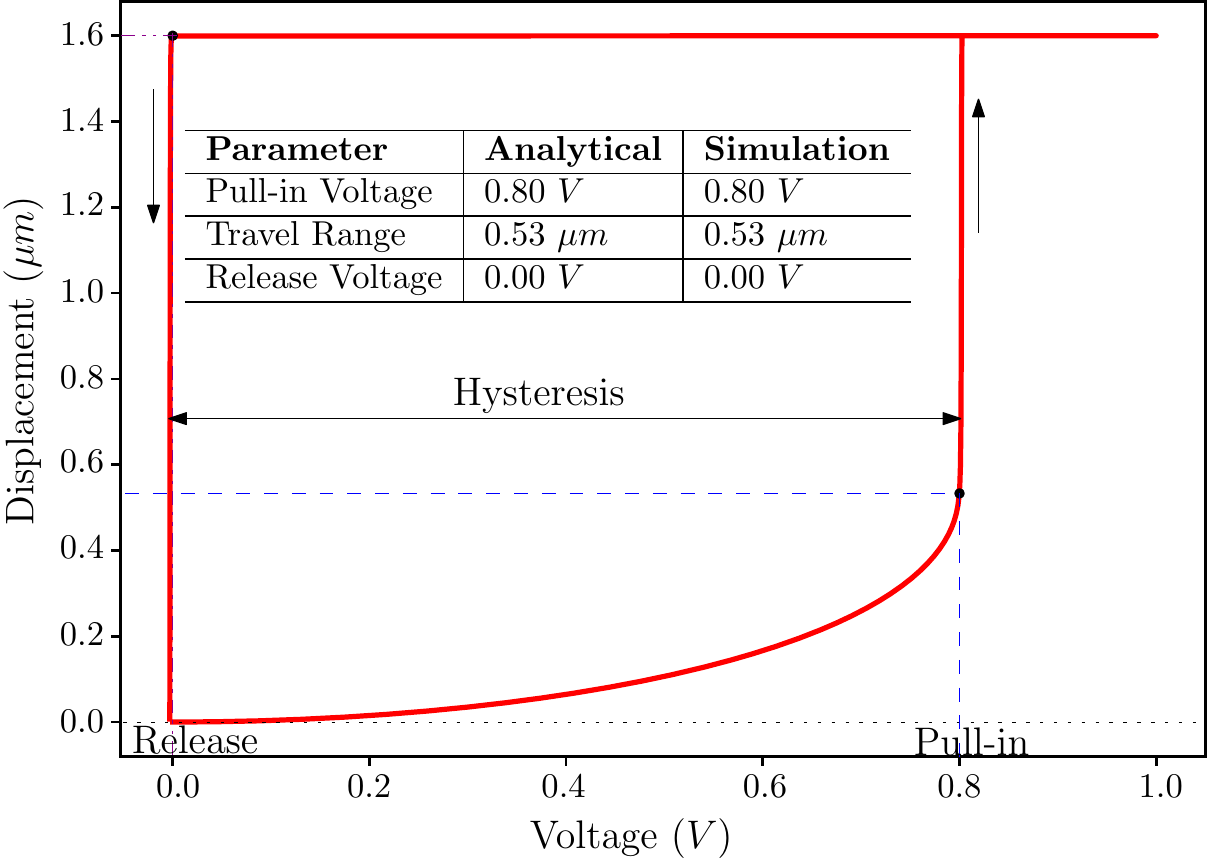}
    \caption{Static pull-in and release  characteristics of the hybrid
actuator. The simulation results are  in agreement with the analytical
results   given   in   Ref.   \textcolor{blue}{\onlinecite{masuduzzaman2014effective}},   thus
validating the hybrid actuator SPICE model.}
    \label{fig:Static_NC_MEMS}
\end{figure}
Assuming zero release voltage, we  have $g_o - h_s = (r_{\alpha
N}/r_{\beta N})~g_o$. Thickness  $t_F$ and area $A_F$  are designed so
as  to obtain  $V_{HSPI}  = 0.80  ~V$. The  static  pull-in and  release
characteristics of the  hybrid actuator are shown in Fig.  \textcolor{blue}{\ref{fig:Static_NC_MEMS}}. The simulation
results  are  in  agreement  with  the  analytical  results  given  in
Ref.  \textcolor{blue}{\onlinecite{masuduzzaman2014effective}},  thus  validating  the  hybrid actuator SPICE model.

\nocite{*}
\bibliography{JAP}

\end{document}